# Prioritizing Technical Debt in Database Normalization Using Portfolio Theory and Data Quality Metrics

Mashel Albarak
School of Computer Science
University of Birmingham, UK
King Saud University, KSA
mxa657@cs.bham.ac.uk

Rami Bahsoon
School of Computer Science
University of Birmingham
UK
r.bahsoon@cs.bham.ac.uk

## ABSTRACT

Database normalization is the one of main principles for designing relational databases. The benefits of normalization can be observed through improving data quality and performance, among the other qualities. We explore a new context of technical debt manifestation, which is linked to ill-normalized databases. This debt can have long-term impact causing systematic degradation of database qualities. Such degradation can be liken to accumulated interest on a debt. We claim that debts are likely to materialize for tables below the fourth normal form. Practically, achieving fourth normal form for all the tables in the database is a costly and idealistic exercise. Therefore, we propose a pragmatic approach to prioritize tables that should be normalized to the fourth normal form based on the metaphoric debt and interest of the ill-normalized tables, observed on data quality and performance. For data quality, tables are prioritized using the risk of data inconsistency metric. Unlike data quality, a suitable metric to estimate the impact of weakly or un-normalized tables on performance is not available. We estimate performance degradation and its costs using Input\Output (I\O) cost of the operations performed on the tables and we propose a model to estimate this cost for each table. We make use of Modern Portfolio Theory to prioritize tables that should be normalized based on the estimated I\O cost and the likely risk of cost accumulation in the future. To evaluate our methods, we use a case study from Microsoft, AdventureWorks. The results show that our methods can be effective in reducing normalization debt and improving the quality of the database.

• **Information systems** → **Relational Database Model;**

## 1 INTRODUCTION

Information systems evolve in response to data growth, improving quality, and changes in users' requirements. The evolution process face many risks, such as cost overruns, schedule delays and increasing chances of failure; it is believed that 80% of the cost is spent on system's evolution [6]. Databases are the core of information systems; evolving databases schemas through refactoring is common practice for improving data qualities and performance, among other structural and behavioral qualities. Database normalization is one of the main principles for relational database design, invented by the Turing Award winner Ted Codd [9]. The concept of normalization was developed to organize data in "relations" or tables following specific rules to reduce data redundancy, and consequently, improve data consistency by decreasing anomalies. The benefits of normalization go beyond data quality, and can have ramifications on improving maintainability, scalability and performance [11], [20]. However, developers tend to overlook normalization due to time and expertise it requires, and instead turn to other strategies such as, creating more indexes or writing extra code fixes to achieve quick benefits. With the growth of data, ad-hoc fixes can become ineffective, calling for costly and inevitable future refactoring to address these pitfalls.

The technical debt metaphor was coined to describe and quantify issues arising from ill, inadequate or suboptimal practices in developing, maintaining and evolving systems, while compromising long-term qualities for short-term benefits, such as fast delivery and immediate savings in cost and effort [10]. The technical debt metaphor can be a useful tool for justifying the value of database normalization, through capturing the likely value of normalization (e.g. quality improvements) relative to the cost and effort of embarking on this exercise, or delaying it. Though technical debt research has extensively looked at code and architectural debts [22], [3], technical debt linked to database normalization, has not been explored, which is the goal of this study. In our previous work [2], we defined database design debt as:

> " The immature or suboptimal database design decisions that lag behind the optimal/desirable ones, that manifest themselves into future structural or behavioral problems, making changes inevitable and more expensive to carry out".

In [2], we developed a taxonomy that classified different types of debts, which relate to the conceptual, logical and physical design of the database.

In this study, we explore a specific type of database design debt that relates to the fundamentals of normalization theory. The underlying assumption of the theory is that the database tables should be normalized to the ideal normal form, a hypothetical fourth normal form, to achieve benefits [14]. While this assumption holds in theory, practically it fails due to the required time and expertise. Adhering to this assumption, we link database normalization debt to tables in the database that are below the fourth normal form.

Given the high cost of database normalization, prioritizing tables to be normalized is challenging. Studies have showed that the most critical technical debt items are the ones whose metaphoric interests grow fast over time [5]. Therefore, a method that would aid database developers and stakeholders to analyze and estimate the impact of normalization debts is highly needed to avoid negative and costly consequences on database qualities due to lack of normalization. Performance and data quality are among the most concerned qualities when it comes to database normalization [11], [24], [32]. In this study, we will analyze the impact of normalization debt on both data quality and performance and use the results to prioritize tables that should be normalized to the fourth normal form.

To measure the impact of normalization debt on data quality, we use the risk of data inconsistency metric of the International Standardization Organization [19]. Unlike data quality, a suitable metric to measure the impact of normalization debt on performance is not available. We analyze database performance, by looking at the Input and Output " I\O " cost of the operations performed on tables below fourth normal form. I\O cost is the number of disk pages read or written to execute the operation [17]. Weakly or un-normalized tables will acquire more pages due to big amount of data duplication they hold and big records size. Therefore, operations executed on those tables will increase the I\O cost, which in turn will affect performance. As part of our contribution, we propose a model to control the debt impact on the I\O cost. We build on Modern Portfolio Theory [25] to manage the debt by prioritizing tables that should be normalized based on the I\O cost and the risk of table's size growth.

The novel contributions of this paper can be summarized as follows:

- Explore technical debt related to database normalization: Define and estimate the impact of normalization debt, where we specifically look at the ramifications of the normalization decisions on performance and data qualities.
- In large-scale systems, designers are often faced with the challenge of limited resources when it comes to normalization and managing the debt. To address this challenge, our approach prioritize tables that should be normalized based on their impact on data quality and performance. For performance, we utilize portfolio theory to prioritize tables that should be normalized, taking into account: debt impact on operations performance and its accumulation over time.

The contribution provides a basis for software engineers and database developers to understand normalization, not only from the technical perspective but also from its connection to debt. Coining normalization with debt hopes to provide a systematic procedure that can eliminate unnecessary normalization efforts, justify the essential ones, and/or prioritize investments in normalization, when resources and budget are limited.

The methods are evaluated using a case study from Microsoft, AdventureWorks database [1], and StoreFront web application built on top of the database [31]. The database has a total of 64 tables, each populated with large amount of data to support e-commerce transactions of a hypothetical cycles retail company.

## 2 BACKGROUND AND MOTIVATION

In this section, we will summarize the key concept of our work. The normalization process was first introduced by Codd in 1970 [9], as a process of organizing the data in relations or "tables". The main goal of normalization is to reduce data redundancy, which is accomplished by decomposing a table into several tables after examining the dependencies between attributes. Benefits of normalization was discussed and proved in the literature [11], [14]. Examples of such benefits include: improving data quality as it reduces redundancy; minimizing update anomalies and facilitating maintenance. Fig 1 illustrates the normal forms hierarchy. Higher level of normal form indicates a better design [14], since higher levels reduce more redundant data. The main condition to go higher in the hierarchy is based on the constraint between two sets of attributes in a table, which is referred to as dependency relationship.

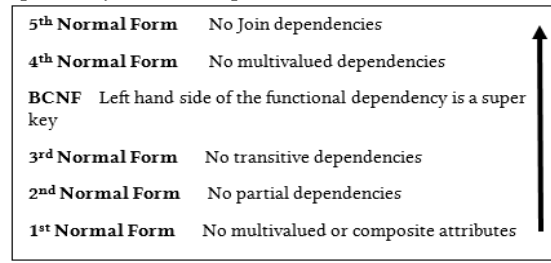

**Figure 1** Normal forms hierarchy

### 2.1 Benefits of Normalization: Data Quality and Performance

Improving data quality is one of the main advantages of normalization [11], [14]. This improvement is linked to decreasing the amount of data duplication as we move higher in the normalization hierarchy. In the presence of data duplication, in poorly or un-normalized tables, there is always a risk of updating only some occurrences of the data, which will affect the data consistency. Data quality is a crucial requirement in all information systems, as the success of any system relies on the reliability of the data retrieved from the system. The trustworthiness of business processes depends highly on the quality of the exchanged data.

Benefits of normalization go beyond data quality and can be observed on other aspects of the system such as performance. Performance has always been a controversial subject when it comes to normalizing databases. Some may argue that normalizing the database involves decomposing a single table into more tables, henceforth, data retrieval can be less efficient since it requires joining more tables as opposed to retrieving the data from a single table. Indeed, de-normalization was discussed as the process of "downgrading" table's design to lower normal forms and have limited number of big tables to avoid joining tables when retrieving the required data [11]. Advocates of de-normalization argue that the Database Management System (DBMS) stores each table physically to a file that maintains the records contiguously, and therefore retrieving data from more than one table will require a lot of I\O. However, this argument might not be correct: even though there will be more tables after normalization, joining the tables will be faster and more efficient because the sets will be smaller and the queries will be less complicated compared to the de-normalized design [11]. Moreover, weakly or un-normalized tables will be stored in a large number of files as opposed to the normalized design due to big amount of data redundancy, and consequently, increased records size and increased I\O cost [24], [11]. Adding to this, not all DBMS store each table in a dedicated physical file. Therefore, several tables might be stored in a single file due to reduced table size after normalization, which means less I\O and improved performance, as proven in [24], [32], [11].

Despite the controversy about normalization and performance, several arguments in favor of normalization is presented by C. J. Date [11], an expert who was involved with Codd in the relational database theory:

- Some of the de-normalization strategies to improve performance proposed in the literature are not "de-normalizing" since they do not increase data redundancy. In fact, as Date states, some of them are considered to be good normalized relational databases.
- There is no theoretical evidence that de-normalizing tables will improve performance. Therefore it's application dependent and may work for some applications. Nevertheless, this does not imply that a highly normalized database will not perform better.

In this study, we view these deteriorations in performance and data quality as debt that can be rooted in inadequate normalization of database tables. The debt can accumulate overtime with data growth and increased data duplication. It is evident that the short-term savings from not normalizing the table can have long-term consequences which may call for inevitable costly maintenance, fixes and/or replacement of the database. Therefore, we motivate rethinking database normalization from the debt perspective linked to performance and data quality issues. Rethinking databases normalization from the technical debt angle can lead to software that has the potential to better evolve and cope with changes and continuous data growth.

## 2.2 Cost of Normalization

As mentioned previously the process of normalizing the database is relatively complex due to expertise and resources required. According to [4], decomposing a single table for normalization involves several refactoring tasks:

- Modifications to the database schema: create the new tables after decomposition; create triggers to ensure that the data in new tables will be synchronized and finally updating all stored views, procedures and functions that access the original table.
- Migrating data to new table/s: a strategy should be set to migrate the data from the old weakly or un-normalized table to the new tables.
- Modifications to the accessed application/s, which involve introducing the new tables' meta data and refactor the presentation layer accordingly.

Additionally, testing the database and the applications before deployment can be expensive and time consuming. Consequently, the aim should not be normalizing all the tables, rather than prioritize tables that has the most negative effect to be normalized. In this study, we aim to formulate the normalization problem as a technical debt. We start from the intuitive assumptions that tables, which are weakly/not normalized to the deemed ideal form, can potentially carry debt. To manage the debt, we adhere to the logical practice in paying the debt with the highest negative impact first. We use the impact information of the debt item to inform normalization decisions, constrained to time, budget and quality objectives.

## 2.3 Database Refactoring

Database refactoring, or some may refer to as schema evolution, has been largely studied over the past years [4], [8], [27]. Schema evolution is basically altering the schema of the database to adapt it to a certain required change[4]. Attempts have been made to analyze the schema evolution from earlier versions [8], and tools have been developed to automate the evolution process [27]. However, schema evolution literature has focused on limited database refactoring tactics, such as adding or renaming a column, changing the data type, etc. Evolving the schema for the purpose of normalizing the database to create value and avoid technical debt has not been explored. We posit that normalization is an exercise that should create a value. By creating a value, it is ensured that the system will sustain and be much more usable and maintainable.

The process of normalizing a database can be very costly since it involves decomposing weakly normalized tables into several tables following specific rules, and update the applications that use the database. Researchers have provided several algorithms to automate database normalization [13], [12]. Their studies aimed to produce up to 3rd normal form or BCNF tables automatically. However, these studies looked at the schema in isolation from applications using the database. The consideration of the applications is important to better estimate the cost of normalization taking into account refactoring, configuration and data migration tasks. Since this process can be costly, our study aims to provide a method to prioritize tables to be normalized to improve the design and avoid negative consequences.

## 3 NORMALIZATION DEBT DEFINITION

There is no general agreement on the definition of technical debt in the literature [21]. However, the common ground of existing definitions is that the debt can be attributed to poor and suboptimal engineering decisions that may carry immediate benefits, but are not well geared for long-term benefits. Database normalization had been proven to reduce data redundancy, improve data quality and performance as the table is further normalized to higher normal forms [11]. Consequently, tables below the 4th normal form can be subjected to debts as they potentially lag behind the optimal, where debt can be observed on data consistency and performance degradation as the database grows [11], [24]. To address this phenomenon of normalization and technical debt, we have chosen 4th normal form as our target normal form. Although most practical database tables are in 3rd (rarely they reach BCNF) normal form [14], 4th normal form is considered to be the optimal target since more duplicate data is removed. Additionally, 4th normal form is based on multi valued dependency which is common to address in real world [34]. Moreover, higher normal form of the 5th level is based on join dependencies that rarely arise in practice, which makes it more of a theoretical value [11], [14].

Following similar ethos of database design debt definition [2], we view that database normalization debt is likely to materialize for any table in the database that is below the 4th normal form.

Using available data in the AdventureWorks data dictionary [1], we have identified 23 debt tables below 4th normal form. Due to space limitation, we will analyze five tables presented below in Table 1:

**Table 1 Debt tables below 4th normal form**

| Table Name | Normal Form |
|---|---|
| Product | Un-normalized |
| Employee | 1st normal form |
| EmployeePayHistory | 1st normal form |
| ProductProductPhoto | 1st normal Form |
| WorkOrder | 2nd normal form |

## 4 NORMALIZATION DEBT IMPACT

Data quality and performance are among the qualities that are affected by database normalization or its absence. Normalization technical debt can be observed on these qualities. As it can be difficult to eliminate the debt, we hope to manage the debt by prioritizing debt tables below the 4th normal form that are believed to have the highest negative effect on them. Henceforth, the first step is to estimate the impact of each debt table on those qualities, where the estimation will be the main driver of prioritizing debts to be paid.

In technical debt area, the debt interest is a crucial factor for managing debts. In general, the interest of technical debt is the cost paid overtime by not resolving the debt [16]. Unlike finance, interest on technical debt has been acknowledged to be difficult to elicit and measure for software engineering problems [5]. The interest can span various dimensions; it can be observed on technical, economic, and/or sociotechnical dimensions, etc. One important aspect of technical debt interest is that it is context dependent, meaning the same detected debt can have more or few interest in the future depending on the project and its circumstances. Researchers have coined interest with implication on qualities [35]. The analogy is applicable to the case of normalization as striving for the ideal 4th normal form will reduce data duplication substantially and consequently, decrease the rate at which the current debt tables negatively affect data quality and performance.

### 4.1 Data quality

Data quality is considered as one of the main advantages of normalization. In this study, data quality is represented by its consistency. In weakly or un-normalized tables that store big amount of data duplication, there is always a possibility to change or update some occurrences of the data leaving the same duplicated data un-updated. Therefore, the risk of data inconsistency will be higher. This risk is decreased by normalization as the amount of data duplication is decreased. Thus, we view the risk of data inconsistency as an impact incurred due to weakly normalized tables. This Impact can be quantified using the International Standardization Organization (ISO) metric [19]. According to ISO, The risk of data inconsistency is proportional to the number of duplicate values in the table and it is decreased if the table further normalized to higher normal forms. This risk can be measured using the following formula:

$X = A/B$

With $\binom{n}{k}$ sets of k attributes for a table with n attributes (k=1,..n), where $A = \Sigma k \Sigma i D i$, $Di$ = number of duplicate values found in set i of k attributes, and B= number of rows × number of columns.

Rows refer to the data stored in the table and columns refer to the attributes of that table. For X, lower is better. Using this metric, developers can prioritize tables to be normalized by calculating the risk of data inconsistency for all the tables in the database and identify tables with higher risks to be normalized.

We have calculated the risk of data inconsistency for the tables in AdventureWorks database, the results are shown in Table 2.

**Table 2: Risk of data inconsistency for the debt tables**

| Table Name | Risk of Data Inconsistency |
|---|---|
| Product | 17 |
| Employee | 8.213 |
| EmployeePayHistory | 1.034 |
| ProductProductPhoto | 0.978 |
| WorkOrder | 2.910 |

Based on the results from the previous table, the logical approach would be to normalize table `Product` to the 4th normal form as it has the highest risk of data inconsistency. It is also observable from the results that the risk of data inconsistency is independent from the normal form of the debt table. As shown, table `WorkOrder` has a higher risk of data inconsistency than tables `EmployeePayHistory` and `ProductProductPhoto` even though they are in a weaker normal form, which denotes that the normal form alone is not sufficient to make the right decision on which table to normalize.

## 4.2 Performance

Enhanced operations performance is one of the advantages of database normalization, as the database will store smaller sets of data in tables due to the decreased amount of data duplication. Operations performed on the database include update; insert; delete data stored in the tables, in addition to data retrieval operations. Each of these operations incur a cost on the number of disk pages read or written, which is referred to as the Input\Output "I\O" cost [17]. The impact of normalization debt can be observed through the I\O cost incurred by the operations performed on the debt table in its current normal form. Due to the huge amount of data redundancy in debt tables below 4th normal form, tables will require more pages to be stored in, which will affect the performance of the operations executed on those tables. Meaning, the more data stored in the debt table, the more disk pages it requires and henceforth, more time to go through the pages to execute the operation. Therefore, normalization is the sensible solution that will not only improve performance, as it has been proven in several studies [24], [32], [11], but it will also improve the overall design of the database to adapt easily to future changes.

Unlike data consistency, a metric to quantify the I\O cost of all the operations performed on debt tables is not available. In the following section we propose a model to estimate this cost and suggest an approach, which is grounded on portfolio approach to manage it effectively.

### 4.2.1 Performance Estimate: The Impact of Normalization Debt on the I\O cost

I\O cost of each operation can be elicited from the database management system via query analyzer tools. To quantify the total I\O cost for each debt table, we use the execution rate of the operation and the I\O cost of operations. Let $n_u$, $n_i$, $n_d$, $n_s$ be the number of update, insert, delete and select operations on the debt table R respectively. $\lambda_x^u$, $\lambda_x^i$, $\lambda_x^d$, $\lambda_x^s$ represent the execution rates of the xth update, xth insert, xth delete and xth select respectively. Finally, the I\O costs of the xth update, xth insert, xth delete and xth select are denoted by $C_x^u$, $C_x^i$, $C_x^d$, $C_x^s$ respectively. It is important to note that operations on a specific single table are considered (excluding the join operations when calculating the impact). The reason for this is because we want to examine the cost incurred by a specific debt table due to its weakly normalized design. Moreover, after the decision is made whether to pay or keep the debt, the join operations are still needed to retrieve the required data. Therefore, the I\O cost of a debt table can be calculated as follows:

$$I\backslash O \text{ cost} = \sum_{x=1}^{n_u} C_x^u \lambda_x^u + \sum_{x=1}^{n_i} C_x^i \lambda_x^i + \sum_{x=1}^{n_d} C_x^d \lambda_x^d + \sum_{x=1}^{n_s} C_x^s \lambda_x^s$$

Let us consider a table named Staff as an example. This table stores staffs' information. Assuming table Staff is in the 2nd normal form, where the ideal is the 4th normal form and henceforth, considered as a debt table. Suppose that there are 2 select statements to retrieve data from this table, 1 update statement to update some information, 1 insert statement and no delete statements. Values of the other variables are assumed as the following Table 3:

**Table 3: Values of table Staff I\O cost variables**

| Variable | Value |
| --- | --- |
| $n_u$ | 1 |
| $n_i$ | 1 |
| $n_s$ | 2 |
| $C_1^u$ | 2 I\Os |
| $\lambda_1^u$ | 100/month |
| $C_1^i$ | 1 I\O |
| $\lambda_1^i$ | 50/month |
| $C_1^s$ | 5 I\Os |
| $\lambda_1^s$ | 200/month |
| $C_2^s$ | 3 I\Os |
| $\lambda_2^s$ | 500/month |

Referring to the previous Table 3, the I\O cost of table Staff's operations would be:

$(2 \times 100) + (1 \times 50) + (5 \times 200) + (3 \times 500) = 2750$ Estimated average I\O cost monthly.

### 4.2.2 Technical Debt Impact Accumulation

Technical debt grows with the accumulated interests on the debt [16]. When likely interest accumulation is overlooked, it can lead to ill-justified decisions regarding debt payment. For example, suppose that that a debt is identified in a software artifact and the developer demands that it should be fixed before the next release. As a result, the developer may end up wasting effort, time and cost for fixing something that may not have severe or noticeable impact on the system or business in the future. In normalization debt context, similar situation is resembled in I\O cost accumulation. I\O cost changes based on the debt tables' growth rate. If the table is likely to grow faster than other tables, the I\O cost for the operations executed on that table will accumulate faster than others. This due to the fact increasing table size implies more disk pages to store the table and therefore, more I\O cost. Tables' growth rate is a crucial measure to prioritize tables needed to be normalized. If the table is not likely to grow or its growing rate is less than other tables, a strategic decision would be to keep the debt and defer its payment. Table growth rate can be elicited from the database monitoring system. The growth rate of a table can be viewed as analogous to interest risk or interest probability. Interest probability captures the uncertainty of interest growth in the future [23]. Debt tables which experience high growth rate in data can be deemed to have higher interest rate. Consequently, these tables are likely to accumulate interest faster.

### 4.2.3 Managing I\O cost of normalization debt (portfolio approach)

#### *4.2.3.1 Modern Portfolio Theory*

Modern Portfolio Theory (MPT) was developed by the Nobel Prize winner Markowits [25]. The aim of this theory is to develop a systematic procedure to support decision making process of selecting capital of a portfolio consisting of various investment assets. The assets may include stocks, bonds, real estate, and other financial products on the market that can produce a return through investment. The objective of the portfolio theory is to select the combination of assets using a formal mathematical procedure that can maximize the return while minimizing the risk associated with every asset. Portfolio management involves determining the types that should be invested or divested and how much should be invested in each asset. This process draws on similarity with the normalization debt management process, where developers can make decisions about prioritizing investments in normalization, based on which technical debt items should be paid, ignored, or can further wait. With the involvement of uncertainty, assets expected return and variance of the return are used to evaluate the portfolio performance. The expected return of a portfolio is presented by the following equation as the weighted sum of the expected return of the assets in the portfolio. The weight of an asset represents the proportion from the capital invested in this asset.

$$E(R) = \sum_{i=1}^{n} X_i E(R_i)$$

Where E(R) is the expected return of a portfolio of n assets, Xi is the weight of asset i. The weight determines the proportion of the money that should be invested in asset i. $E(R_i)$ is the expected return of that asset. This expected return of the portfolio is constrained by the following equation, which denotes the sum weight of all the assets in the portfolio should equal to one:

$$\sum_{i=1}^{n} X_i = 1$$

To measure the risk of a financial portfolio, the variance and standard deviation is used. The risk of a portfolio depends on:

- The volatility of each asset's return which is estimated based on the observation of its return over time.
- The weight invested in each asset
- The correlation between assets return which is estimated by the observation of assets overtime.

The risk of a portfolio R is calculated using the following equation [25]:

$$R = \sqrt{\sum_{i=1}^{n} X_i^2 \sigma i^2 + 2 \sum_{i<j}^{n} X_i^2 p_{ij} \sigma i^2 \sigma j^2}$$

Where Xi is the weight of an asset i, σ is the variance of this asset return and $P_{ij}$ is the correlation between asset i and asset j.

#### *4.2.3.2 Modelling the problem (Portfolio Based Approach to Manage the I\O cost of Normalization Debt)*

Our approach aims to help database developers make strategic decisions about refactoring the database for normalization. Let us consider an existing database system with tens of tables, all or most of which are below the 4th normal form. Pervious research and the classical theory of normalization would encourage normalizing all tables to the ideal normal form [11]; the exercise would require a lot of time and resources, where a complete refactoring of the whole system might be an alternative cost saving option. Our approach acknowledges the fact that time, resources, and budget is often a constraint that prevent exhaustive and unjustified normalization. Our approach selects and prioritizes the tables to be normalized by constructing a portfolio of multiple debt tables that has the highest priority for normalization. The objective is to minimize the negative impact of the debt tables on performance taking into consideration the likely growth rate of the table size and henceforth, the risk of interest accumulation.

We view a database of debt tables below the 4th normal form as a market of assets. To fit in portfolio management, each debt table is treated as an asset. For each table, we need to determine whether it is better to normalize that table to the 4th normal form (pay the debt) or keep the table in it is current normal form ( defer the payment). To decide on this, we need to determine what the expected return of each debt table is. In the case of normalization debt, the expected return of the debt table resembles the estimated performance impact of the table. Tables with the lowest estimated impact are deemed to carry higher expected return. In other words, If the estimated I\O of table's A operations is less than estimated I\O of table's B operations, then table A expected return would be higher than B; B will then has a higher priority for normalization due to high I\O. We balance the expected return with the risk. In portfolio management, this risk is represented by the variance of the return. For the debt tables, this risk is represented by the tables' growth rate. Tables with the highest growth rate are considered to be risky assets, their likely interest and so the debt will grow faster than other tables of low growth rate.

In order to apply the portfolio theory to normalization debt, few considerations need to be taken into account:

- The expected return of the debt table is equal to 1/I\O cost.
- The risk of each debt table is equal to the table growth rate for each debt table. This information can be elicited from the database management system by monitoring the table's growth.
- We set the correlation between the debt tables to zero for several reasons: First, the I\O costs of the debt tables are independent. Meaning, the I\O cost of the operations executed on a debt table has no effect on the I\O cost of the operations executed on another debt table. Moreover, the growth rate for each table, which affects the I\O cost, is unique and independent from each other. Lastly, each debt table design is independent from other debt tables, as the decision to keep the debt or normalize the table have no effect on the design and the data of the other debt tables.

Taking into account these considerations, we can apply the portfolio theory, where the database developer is investing in tables' normalization. The database developer needs to build a diversify portfolio of multiple debt tables. Multiple debt tables in the database represent the assets. For each asset i, it has its own risk $R_i$ and I\O cost $C_i$. Based on these values the developer then can prioritize tables to be normalized. The expected return of debt tables portfolio Ep, built by prioritizing debt tables from the database of m debt tables can be calculated as in the following equation:

$$Ep = \sum_{i=1}^{m} w_i \frac{1}{C_i}$$

With one constraint represented in the following equation:

$$\sum_{i=1}^{m} w_i = 1$$

Where wi represents the resulted weight of each debt table. This weight will resemble the priority of each table for normalization as explained in the process steps.

The risk of table growth rate for debt table i is represented by Ri. The global risk of the portfolio Rp is calculated as the following:

$$Rp = \sqrt{\sum_{i=1}^{m} w_i^2 R_i^2}$$

*Process Steps:*

1. Identify debt tables: tables below 4th normal form should be identified. If this is not already documented, it would require knowledge of the functional dependencies and rules which can be elicited from the requirements and business analysts.
2. Determine the debt tables' growth rate from the database monitor. This step will simplify the method to examine only debt tables with high growth rate.
3. Consider only the tables with high growth rates to calculate I\O costs of their operations.
4. For each debt table of high growth rate, list all queries, update, insert and delete operations, execution rate for each operation and their I\O costs
5. Calculate the I\O cost of each table's operations as explained in section 4.2.1
6. Determine the values of the portfolio model variables (expected return= 1/I\O cost) and (risk= table growth rate)
7. Run the model on the available data to produce the optimal portfolio of the debt tables. The portfolio model will provide the highest weights to those tables with low I\O cost and low table growth rate. Therefore, debt table that has the lowest weight implies the highest priority table that should be normalized.
8. Use the results to justify the decisions to stakeholders.

It is important to mention that this process should be executed iteratively before each release for both the I\O cost and growth rate of debt tables vary during the system's life.

**4.2.4 Case study**

We considered the AdventureWorks database, designed by Microsoft [1] and StoreFront web application built on top of the database [31].This database supports standard e-commerce transactions for a fictitious bicycle manufacturer. The database has a total of 64 tables, each filled with thousands of fictitious data. The data dictionary is available with a fair description of the tables and the attributes, which will facilitate the process of identifying the normal form for each table. To better understand our approach, the following example demonstrates how the steps performed to manage the impact of normalization debt on performance.

**Step 1:** Identify debt tables: 23 tables have been identified below 4th normal and considered to be debt tables.

**Step 2 & 3:** Growth rate can be monitored or retrospectively captured from the database monitoring system. For simplicity, we assume that this procedure has identified the following 5 tables, presented in Table 4 to have the highest growth rate among the debt tables.

**Table 4: Debt tables with the highest growth rate**

| Table name | Growth rate\monthly |
|---|---|
| Product | 0.2 |
| Employee | 0.1 |
| EmployeePayHistory | 0.1 |
| ProductProductPhoto | 0.2 |
| WorkOrder | 0.5 |

**Step 4:** A list for each table has been constructed similar to the following Table 5 with all the select, insert, update delete statements, their assumed execution rates and I\O cost per execution. The following Table 5 demonstrates a list of table `Product` operations:

**Table 5: Table Product list of operations**

| Select Statements | Execution rate per month | I\O cost per execution |
|---|---|---|
| 1 | 3000/month | 2 |
| 2 | 5000/month | 2 |
| 3 | 4000/month | 1 |
| Update statements | Execution rate per month | I\O cost per execution |
| 1 | 2000/month | 2 |
| 2 | 3000/month | 1 |
| Insert statements | Execution rate per month | I\O cost per execution |
| 1 | 100/month | 1 |

**Step 5:** The following Table 6 shows the calculated I\O cost of the operations performed on each debt table as detailed in section 4.2.1

**Table 6: I\O cost of the debt tables**

| Table name | I\O cost |
|---|---|
| Product | 27100 |
| Employee | 15000 |
| EmployeePayHistory | 12000 |
| ProductProductPhoto | 20000 |
| WorkOrder | 4000 |

**Step 6:** Table 7 presents the values of the portfolio model variables (expected return and risk); the expected return values presented in the table are rounded values.

**Table 7: Portfolio model variables**

| Table name | Expected return=1/I\O cost | Risk=growth Rate/ monthly |
|---|---|---|
| Product | 0.003 | 0.2 |
| Employee | 0.006 | 0.1 |
| EmployeePayHistory | 0.008 | 0.1 |
| ProductProductPhoto | 0.005 | 0.2 |
| WorkOrder | 0.025 | 0.5 |

**Step 7:** After running the portfolio model on the available data, the following weights presented in Table 8 were determined for each debt table:

**Table 8: Debt tables weights**

| Table name | Weight |
|---|---|
| Product | 4.4 |
| Employee | 35.29 |
| EmployeePayHistory | 47.06 |
| ProductProductPhoto | 7.3 |
| WorkOrder | 5.8 |

From Table 8 we can determine that table `Product` has the highest priority to normalize since it got the lowest weight. Although this table's growing rate is considered to be small when compared to table `WorkOrder`, the I\O cost incurred by the operations executed on table `Product` is the highest, meaning the number of operations performed on that table and the frequencies of those operations on a monthly basis are high. Henceforth, this table will have more I\O cost compared to the other tables; this I\O cost will accumulate faster than other tables despite the relatively smaller table growth rate. On the other hand, table `WorkOrder` has the second priority for normalization despite its lowest I\O cost. This is because it has the fastest growth rate among the tables, which will accelerate I\O cost accumulation in the future and affect performance. As seen, the difference of the weights between table `Product` and table `WorkOrder` is relatively small, which indicates that both tables are semi-equally important to normalize, considering time and budget constraints. Tables `Employee` and `EmployeePay History` are the tables with the least priority to be normalized since both their I\O costs and growth rates are the smallest among tables. However, if time and budget permits only one table to normalize, table `Product` would be the correct choice since it also has the highest risk of data inconsistency. Therefore, by normalizing this table, both performance and data quality will be improved.

## 5 EVALUATION

One objective of the case study is to investigate how the described methods can be applied by database developers to reason about normalization decisions; improve data quality and performance in an attempt to manage normalization debts and their accumulated interests, while minimizing the cost of database refactoring. The evaluation contrasts our debt-aware approach to the ad-hoc one. The debt metric provides insights on the significance of the impact of weakly or un-normalized tables and their accumulation leading to deterioration in the quality. We have claimed the 4[th] normal form as the target normal form for the "debt-friendly" table design. However, practically, achieving this target for all tables, as conventional approaches of database normalization suggest, is idealistic and costly process due to constraints on expertise, time and budget. Therefore, our approach facilitates the decision making process for normalization debt management, regarding which table has a higher priority to normalize, taking into consideration the effect of the debt table and likely future quality degradations overtime. In particular, our debt-aware approach provides insights on whether it is beneficial to go for the 4[th] normal form based on likely benefits relative to improving the data quality and performance due to this exercise. Table 9 presents how our prioritization methods has followed a systematic and informed procedure in significantly reducing normalization effort while improving qualities (rather than ad-hocly targeting all the tables that are below 4[th] normal form).

**Table 9: Difference in effort between conventional approach and debt-aware approach of normalization**

| Approach | Number of tables to normalize |
|---|---|
| Conventional approach | 5 |
| Prioritize based on risk of data inconsistency | 1 |
| Prioritize based on I\O cost | 2 |

Table 9 shows that following the conventional approach, which encourage normalizing all the tables to 4$^{th}$ normal form is more costly; time consuming and ad-hoc than the debt-aware approach for two scenarios: The first scenario prioritizes debt-tables based on their impact on data quality, measured using the ISO metric Risk of Data Inconsistency [19], where it suggests to only normalize one table, table `Product`, based on the biggest amount of data duplication it holds. The second scenario considers the performance impact of each debt table, and the likely accumulation of this impact in the future. By utilizing the portfolio approach, two tables are suggested for normalization, `Product` and `WorkOrder`. Moreover, depending on available resources, developers can include more tables to normalize and justify their decisions based on the debt tables' impact on data quality and performance. In summary, our debt-aware approach has provided more systematic and informed cost-effective decision than the conventional approach for normalization taking into consideration debts linked to performance and data quality.

Though our approach has looked at two essential qualities in database normalization, performance and data quality, the approach can be extended and be applicable to reason about the likely debts and interests relative to other structural or behavioral qualities such as maintenance, availability among the others. The extension requires identifying appropriate metrics for the analysis, however. The inclusion of other qualities has the promise to provide database designers and developers with more comprehensive approach for prioritization and debt management.

## 6 RELATED WORK

Technical debt metaphor was first introduced by Ward Cunningham in 1992 [10]. Back then he described coding debt as a trade-off between short term goals (i.e. shipping the application early to meet the market) and applying the optimal coding practices for long term goals. Technical debt has captured the attention of many researchers over the past years. The attempts have exceeded code level debt to encompass other aspects of the system such as architectural debt [23], requirements debt [15], testing debt [29], among other types of technical debt [22], [16]. As our work is closed to debt prioritization, authors in [35] utilized prioritization to manage code level debts in software design. The authors estimated the impact of God classes on software maintainability and correctness. They prioritized classes that should be refactored based on their impact on those qualities. Portfolio Theory was proposed to by researchers to manage technical debt [30], [18]. In [18] the authors viewed each debt item as an asset, and they utilized Portfolio Theory to construct a portfolio of debt items that should be kept based on debt principal, which they defined as the effort required to remove the debt, and debt interest which is the extra work needed if the debt is not removed. Portfolio Theory was also proposed to manage requirements compliance debt in [28]. The authors viewed compliance management as an investment activity that needs decisions to be made about the right compliance goals under uncertainty. They identified an optimal portfolio of obstacles that needed to be resolved, and address value-driven requirements based on their economics and risks. Despite the vast contributions on technical debt in software, database design debts received very little attention. In [33], the authors studied technical debt related to referential integrity constraints in databases schema. They viewed missing foreign keys in the tables as a debt that will affect data quality. They proposed an iterative process to reduce the debt, which involves measuring the debt by making use of missing foreign keys detection algorithms; define modification activities that should be done in the development site and finally, package all modifications to the deployment sites. They illustrated their approach using OSCAR electronic medical record system.

Technical debt interest is among the essential elements needed to manage the debt effectively [7], [26]. Ampatzoglou et al. [5] presented a framework to manage the interest of technical debt. Their framework involved interest definition; classification; evolution and finally, interest management theory, based on Liquidity Preference Theory.

## 7 CONCLUSION

We have explored a new context of technical debt, which is linked to database normalization. Normalizing the database is essential to improve data quality and performance. However, practically, developers tend to overlook normalization process due to time and expertise it requires, and instead, turn to other quick procedures, such as creating more indexes or writing extra code fixes to save time and effort. With data growth and systems' evolution, normalizing the database becomes essential to avoid long-term effects on data quality and performance. We have asserted that normalization debt is likely to be materialized for tables below the 4$^{th}$ normal form and we discussed the validity of this assertion.

Conventional approaches for normalization encourage normalizing all the tables to achieve benefits; the exercise is often judged in isolation of the cost, long-term value and technical debt avoidance. Furthermore, it is impractical to embark on exhaustive normalization of all tables because of costs and uncertain benefits. In this study, we proposed an approach to manage the debt by prioritizing tables that should be normalized. The prioritization is based on the impact of the weakly or un-normalized tables on data quality and performance. To improve data quality, the prioritization is based on a metric provided by the ISO [19] to measure the risk of data inconsistency for each debt table below the 4$^{th}$ normal form. To enhance performance, a model was proposed to estimate the I\O cost of the operations performed on each weakly or un-normalized table. These costs tend to increase with the growth of the tables. To manage this debt effectively and avoid accumulation of I\O cost, Portfolio Theory was utilized to prioritize tables that should be normalized based on the I\O costs of the operations performed on the tables and the risks of future cost accumulation (e.g., interest on the debt). The techniques was applied to AdventureWorks database from Microsoft. The results show that rethinking conventional database normalization from the debt angle can provide more systematic guidance and informed decisions to improve the database quality, while reducing the cost and effort that is linked with unnecessary normalization.